\documentclass[%
 reprint,
 amsmath,amssymb,
 aps,
]{revtex4-1}

\usepackage{graphicx}
\usepackage{dcolumn}
\usepackage{bm}

\begin{document}	
\title{A quantum mechanics picture}%

\author{Jeconias Rocha Guimar\~{a}es}

\affiliation{
 Universidade Tecnol\'{o}gica Federal do Paran\'{a}\\
}

\date{\today}

	\begin{abstract}
A quantum mechanics representation based on position ($\vec{r}$), linear momentum($\vec{p}$) and energy($E$) eigenvalues is presented here. A set of equations, explicitly independent on wave function, was derived relating these observables. In this view, a particle has a known trajectory and at any point on space there is a linear momentum associated. Trajectory here, can be viewed as if a measurement were taken continuously. This picture does not change current quantum mechanics interpretation, rather it comes as a new route of calculation. Also, wave function can be retrieved performing an integration in all space for linear momentum. 
Equations derived in this work, present a potential dependent on linear momentum originating what we call quantum force. A particle experiences a total force whose resultant is composed by classic and quantum force, evolving on time according to it. In terms of evaluation, a single particle could be described by a set of auxiliary particles named as wave particles(WP). They will start with a range of initial conditions and will collectively describe probabilistic aspects of quantum mechanics. We expect this route could be applied to many problems, such as multi-body systems.
	
	\end{abstract}

\maketitle

\section{Introduction}

Success of quantum mechanics(QM) has been extensively demonstrated since its first concepts were developed. However, there is still a debate about wave function interpretation. The Copenhagen interpretation, which is the most accepted, could be regarded as the QM standard view\cite{heisenberg1999physics, wheeler1983quantum}. It states that any particle is in a quantum state that represents the possible results of a measurement. According to this, when a measurement is taken, wave function collapses to a single result. Consecutive measurements will produce different results. Consequences of this interpretation lead to some critics and apparent paradoxes, like the one pointed by Einstein, Podolsky and Rosen \cite{EinsteinPodolskyRosen1935}. The most notable attempt to avoid those consequences is the bohmian mechanics\cite{PhysRev.85.166} that brings determinism back to explanation of quantum particles. This interpretation maintains Schr\"{o}dinger's equation(equation\ref{SchEq}) \cite{Schrodinger1926} at the center of non relativistic quantum mechanics. On this view, wave function produces a quantum potential, or field, that is combined with classic one and in which particles evolve on time in a very complex pattern \cite{PhysRevA.88.022116}. The apparent lack of determinism on measurements would come from the impossibility of knowing the exact initial conditions of particles. Unlike bohmian mechanics me make no use of an \textit{ansatz} on $\psi$, we rely only on time independent Schr\"{o}dinger equation and most importantly, we are not presenting a new interpretation on quantum mechanics, rather all aspects of Copenhagen interpretation are retained. Our view offers a possibility to describe electrons in a new way bringing insights on all these unique behaviors. Also, it is interesting analyze within this perspective some problems that represents a very hard task to be described, like multi-electronic. This class of system is a challenge on quantum mechanics, essentially due the size of wavefunction\cite{CAPELLE2006}. There are some ways to deal with it, we mention mean field approximation (Hartree-Fock)\cite{PhysRev.35.210.2}, Density Functional Theory (DFT)\cite{Hohenberg1964,Kohn1965}, Perturbation theory\cite{PhysRev.46.618}, Configuration of interaction (CI) \cite{DAVIDSHERRILL1999143} and Coupled Cluster\cite{doi:10.1063/1.1727484}. The last three are usually called post Hartree-Fock(HF) methods, since, differently of DFT, are based on wave function. Mean field theories, are built in an approximation where total wave function is written in terms of combination of individuals electronic wave function. Such procedure usually results on poor results for most of systems. Post Hartree-Fock methods make improvements on it, for instance, on CI methodology electrons are promoted to virtual HF levels and total wave function is then written in terms of these configurations. DFT, alternatively, uses electronic density as its main variable mapping many-body problem into a single-electron one. Two Theorems, in which DFT is based, ensure there is a functional of electronic density that allows calculation of ground energy exactly. In analogy to DFT, our method focus on linear momentum as central variable.

\begin{equation}
-\frac{\hbar^2}{2m} \nabla^2\psi + U\psi=E\psi
\label{SchEq}
\end{equation}

\section{Deriving}
\label{sec:Deriving}

First, we will eliminate wave function from our quantum equation. So, lets write partial derivatives of wave function or, equivalently, the definitions of linear momentum operator:

\begin{equation}
\frac{\partial \psi}{\partial x_k}=\frac{i}{\hbar}p_{x_k}\psi
\label{lmom1}
\end{equation}

Equation \ref{lmom1} can be differentiated again:

\begin{equation}
\frac{\partial^2 \psi}{\partial x_k^2}=\frac{i}{\hbar}\frac{\partial p_{x_k}}{\partial x_k}\psi +  \frac{i}{\hbar}p_{x_k} \frac{\partial \psi}{\partial x_k}
\label{d2lmom}
\end{equation}
 
Replacing single partial derivatives found on equation \ref{lmom1} on equation \ref{d2lmom} and summing on all directions we have:

\begin{equation}
\sum_k\frac{\partial^2 \psi}{\partial x_k^2}=\frac{i}{\hbar}\sum_k\frac{\partial p_{x_k}}{\partial x_k}\psi  -  \frac{1}{\hbar^2} \sum_kp^2_{x_k} \psi  
\label{zeta2}
\end{equation}

Laplacian on the left side of equation \ref{zeta2} can be replaced by Schr\"{o}dinger equation(eq. \ref{SchEq}), resulting in an equation for energy that depends only on eigenvalues of linear momentum, then for $\psi \neq 0$:

\begin{equation}
E=\frac{p^2}{2m}+U-i\frac{\hbar}{2m}\nabla \cdot \vec{p}
\label{EnergyP}
\end {equation}

Solving for all directions, equation \ref{lmom1} leads to a relationship between wave function and linear momentum:

\begin{equation}
\psi= e^{ \frac{i}{\hbar}\int \vec{p}\cdot \partial\vec{r} }
\label{wav}
\end{equation}

Thus, all aspects related to probabilistic are obtained by this integration of $p$ on space.

Another differential operator can be found for $p$, such as its curl, which can be achieved by commutation relation between two components of linear momentum operator $\hat{P}_{x_k}$:

\begin{equation}
\hat{P}_{x_k}\hat{P}_{x_j}\psi-\hat{P}_{x_j}\hat{P}_{x_k}\psi=0
\end {equation}

Resulting in the very useful relation below:
\begin{equation}
\nabla \times\vec{p}=0
\label{curlp}
\end {equation}

The main aspect of this picture is given by assumption that we can calculate an expression for force taking the gradient of quantum potential $V$:

\begin{equation}
V=U-i\frac{\hbar}{2m}\nabla \cdot \vec{p}
\label{QV}
\end{equation}

Taking the gradient of $V$ (eq. \ref{QV}) and using free curl of $p$ (eq. \ref{curlp}) we find the following force:

\begin{equation}
\frac{d\vec{p}}{dt} =-\nabla U +i\frac{\hbar}{2m} \nabla^2 \vec{p}
\label{ForceP}
\end {equation} where $\nabla^2 \vec{p}$ is the vector laplacian. Force as obtained above implies on the classic relationship:

\begin{equation}
\vec{p}=m\frac{d\vec{r}}{dt}
\label{pclassic}
\end{equation}

Even though we use \ref{pclassic}, we do not make a semi-classic method. All aspects of quantum mechanics will correctly be described by potential $V$.

Let us apply divergence theorem to equation \ref{EnergyP}:

\begin{equation}
\oint \vec{p}\cdot \hat{n} dS =i\frac{2m}{\hbar}\int\left(E-\left(\frac{p^2}{2m} + U\right) \right)dV
\label{divergTheorem}
\end{equation}

Choosing a cylinder with infinitesimal small lids, height tending to zero and taking $\hat{n}=\hat{x_k}$, we find:

\begin{equation}
p_{x_{k}}(x_k+\delta x_k)=p_{x_{k}}+i\frac{2m}{\hbar}\left(E-\left(\frac{p^2}{2m} + U\right) \right)\delta x_k
\label{px}
\end{equation}

Taking equation \ref{pclassic} and \ref{px}, we find a relationship for time evolution of $p_{x_{k}}$ for a small time displacement:

\begin{equation}
p_{x_{k}}\left(  1- i\frac{2}{\hbar}\left(E-\left(\frac{p^2}{2m} + U\right) \right)\delta t \right)=p_{x_{k}}^0
\label{pxTimesmallEvolv}
\end{equation}

For any displacement, a relationship can be derived as well:

\begin{equation}
p_{x_{k}}(t)=p_{x_{k}}(t_0)e^{i\frac{2}{\hbar}\int_{t_0}^{t} \delta E dt }
\label{pxTimeEvolv}
\end{equation}

with $\delta E=E-\left(\frac{p^2}{2m} + U\right) $. Also, a equation for force can be obtained starting with equation \ref{px}:

\begin{equation}
F_{x_{k}}= i\frac{2p_{x_{k}}}{\hbar}\left(E-\left(\frac{p^2}{2m} + U\right) \right)
\label{ForcPx}
\end{equation}

Equations \ref{ForcPx} and \ref{ForceP} are equivalent and can be used to perform dynamic.

On Following sections we solve Harmonic Oscillator and Hydrogen atom using equations derived above.

\section{Harmonic Oscillator}
\label{sec:HarmonicOscillator}

Writing equation \ref{EnergyP} for an one dimension harmonic oscillator we have:

\begin{equation}
E=\frac{p^2}{2m}+\frac{m\omega^2x^2}{2}-i\frac{\hbar}{2m}\frac{dp}{dx} 
\label{EnergMHS}
\end {equation}

Taking the limit of $x\rightarrow \infty$ and assuming in this limit $\left|\frac{m\omega^2 x^2}{2}\right|>>\left|\frac{\hbar}{2m}\frac{dp}{dx}\right|$ we have asymptotic equation for any $p$ as:

\begin{equation}
p=i m\omega x
\label{EnergMHSinfity}
\end {equation}
In fact this is the solution for $p$ in the ground state. Thus, calculating energy with this solution gives the well known energy $E=\frac{\hbar \omega}{2}$.

Now, if $p$ and its space derivative are very large:

\begin{equation}
p=-i\frac{\hbar}{x}
\label{highp}
\end {equation}

Therefore, we may suggest as solution an equation with the form:
\begin{equation}
p=i\left( C_1 x + \frac{C_2}{x}  \right)
\label{guessSolutions}
\end {equation}

With this solution, we rewrite equation \ref{ForceP} as:

\begin{equation}
\frac{p}{m}\frac{dp}{dx} =-m\omega^2 x +i\frac{\hbar}{2m} \frac{d^2 p}{dx^2}
\label{ForcePx}
\end {equation}

We find constants for our guess in equation \ref{guessSolutions} by placing it on \ref{ForcePx}. Then, the second state for momentum is:

\begin{equation}
p_2 =im\omega x - i\frac{\hbar}{x}
\label{p2MHS}
\end {equation}

Given $p_2$ as found on \ref{p2MHS}, we find dynamics for a quantum particle in second state. Such solution is strongly dependent on initial position, shown below:

\begin{equation}
x(t)=\left( \frac{\hbar}{m\omega}\left( 1-\left(1-\frac{m\omega x_0^2}{\hbar}\right) e^{2i\omega t} \right)  \right)^{1/2}
\label{xtp2}
\end{equation}

Oscillation occurs around the position of highest probability to find a particle($x_M$). As closer $x_0$ is set to $x_M$, more restrict will occurs oscillation. A particle will remain still when initial position is set exactly equal to $x_M$. When it is set to be infinity, classical solution is find. For this case, in terms of Wave Particles, a set of them, not interacting with one another, starting in several initial positions, will have a variable density of population. Initially they could, for instance, be spatially 

Resultant force, comprising classic and quantum, for second state is given by:

\begin{equation}
F=\frac{\hbar^2}{mx^3}-m\omega^2x
\end{equation}

Oscillation is obtained for a particle upon this force around the point where $F=0$, which occurs at $x_M=\sqrt{\hbar/m\omega}$.

\section{Hydrogen atom}
\label{Hatom}

This well known system is composed by one electron hold by one proton due Coulomb potential:

\begin{equation}
U=-\frac{1}{4\pi \epsilon_0}\frac{e^2}{r}
\end{equation}

where $e$ and $\epsilon_0$ are the fundamental electron charge and vacuum permittivity, respectively. In order to write a solution for $p$ we make on equation \ref{EnergyP} $r\rightarrow \infty$ resulting on $U\rightarrow 0$, thus linear momentum is a constant given by $p=\sqrt{2mE}$. Also, in the limit where $U<<\left|\frac{\hbar}{2m}\frac{dp}{dr}\right|$, yields $p=-i\hbar/r$. Combining these solutions and placing it on \ref{EnergyP}, $p$ for first energy level is:

\begin{equation}
p=i\frac{me^2}{4\pi\epsilon_0}-i\frac{\hbar}{r}
\label{pguessHyd}
\end{equation}

Total force acting on electron is now composed of attractive and repulsive terms, with the last one originated purely by quantum mechanics:

\begin{equation}
F=-\frac{1}{4\pi\epsilon_0}\frac{e^2}{r^2}+\frac{\hbar^2}{mr^3}
\label{FHyd}
\end{equation}

Repulsive term goes faster to zero than classic one. Then classic term dominates for $r\rightarrow\infty$. With force \ref{FHyd}, an electron will oscillate around the point where force is zero, which occurs at $r=4\pi\epsilon_0 \hbar^2/e^2$. This point corresponds to highest probability to find electron around nucleus for this atom in ground state.
Force \ref{FHyd} has same direction of vector $\vec{r}_e-\vec{r}_N$, written in terms of electron ($\vec{r}_e$) and nucleus ($\vec{r}_N$) position vetors. From nucleus perspective we must have the same force with opposite sign. Thus, a conservation of linear momentum is kept for this system.

\section{Energy Correction}
\label{ECorrection}
Wave particle will evolve on time by means of force as given by equation \ref{ForcPx}, so we must have a initial guess for linear moment and energy. In order to correct energy, so it will be in a stationary state, we use equation \ref{ForceP} to find energy correction $\nabla E \vec{\delta r}$ as follows:

\begin{equation}
\nabla E=-\frac{d\vec{p}}{dt}-\nabla U +i\frac{\hbar}{2m} \nabla^2 \vec{p}
\label{Ecorr}
\end {equation}

Let us take the case of an electron under Coulomb potential, since we know energy solution, we write classic force as:

\begin{equation}
\vec{F}= -\sqrt{\frac{2\hbar^2 \left| E_0 \right| }{m}}\frac{1}{r^2}\hat{r}
\label{Fenergy}
\end{equation}

Our initial guess for energy is given by $E$, which differs from correct stationary state by $\delta E$. Thus, assuming it is close to correct energy, we may write linear momentum as:

\begin{equation}
p= i\sqrt{2m\left| E \right|}-\frac{i\hbar}{r}
\label{pguess}
\end{equation}

Equation \ref{Ecorr} is then written as:

\begin{equation}
\nabla E = \frac{\hbar}{r^2}\sqrt{\frac{2}{m}}\left( \sqrt{\left| E \right|} - \sqrt{\left| E_0 \right|} \right)\hat{r}
\end{equation}

Taking guess energy as $|E|=|E_0|+|\delta E|$ with $\delta E \rightarrow 0$, correction is given as:

\begin{equation}
\nabla E \delta r = \frac{\hbar}{r^2} \frac{\delta E}{\sqrt{2m|E_0|}} \delta\vec{r}
\end{equation}
 
Term $\delta \vec{r}$ must be seem as a parameter during energy correction. As $r\rightarrow \infty$ correction will be zero, meaning particle is free and any value is accepted. Inversely, energy correction is more effective when electron is closer to maximum probability position($r_M$), maximum correction is found when $r^2=r_M\delta r$. This procedure can be extended to any potential such as one generated by many electrons. 

\section{Conclusion}
The main aspect of this methodology is that it is based on eigenvalues. All aspects from quantum mechanics are retained, since we keep its standard interpretation. Once linear momentum is calculated, wave function can be found, making $p$ to be the core variable here. We derived relations for energy and force, suggesting the last one to be employed in dynamic calculations. Well known problems, found on text books, are solved within this picture, like Harmonic oscillator and Hydrogen atom. When solving Harmonic Oscillator we obtain oscillation of a wave particle, pointing that only a set of them will fully describe this system, since they are placed in a range of initial positions. A procedure for correcting initial guess energy is discussed, where will be null when stationary state is achieved.

Finally, we suggest this method could be employed on challenging problems, such as multi-electronic. Classically, it is possible find a numeric solution using Newton's law, where a set of point particles evolving on time have its electron-electron correlation correctly described. Thus, with this picture we take this advantage of classic dynamics with all quantum aspects retained.



%

\end{document}